# Unique photoluminescence response of $MoS_2$ quantum dots over wide range of As (III) in aqueous media

Jamilur R. Ansari[1], Md. Farhan Naseh[1], Neelam Singh[1], Tapan Sarkar[2],* and Anindya Datta[1],*

[1]*University School of Basic and Applied Sciences, Guru Gobind Singh Indraprastha University, Sector 16-C, Dwarka, New Delhi-110078, India*

[2]*University School of Chemical Technology, Guru Gobind Singh Indraprastha University, Sector 16-C, Dwarka, New Delhi-110078, India*

**Corresponding Authors:**

**Anindya Datta (AD): anindya.datta@ipu.ac.in (email), +91-11-2530-2401 (telephone).**

**Tapan Sarkar (TS): tapan@ipu.ac.in (email), +91-11-2530-2476 (telephone)**

**Abstract**

Solvothermal synthesis of $MoS_2$ based quantum dots (QDs) and the performance evaluation of bare QDs for the detection of aqueous As (III) oxidative state at room temperature and neutral pH over an extremely wide range (0.1 ppb to 1000 ppb) is reported here. Concentration-dependent photoluminescence (PL) of the QDs was found to be enhanced up to 50 ppb and then suppressed till 1000 ppb, showing two distinctive slopes for enhancement and suppression. Passivation of trap states or defects of QDs may be the possible reason for enhancement, and the formation of extremely small glassy $As_2S_3$ particles on the QD surface may be the possible reason for suppression. The pattern of optical absorption of QDs is featureless but shows an enhanced absorbance in the near UV range below $\leq$ 300 nm, which increases with As (III) concentration up to 50 ppb and then decreases following the PL pattern. The $MoS_2$ QDs were characterized by transmission electron microscopy (TEM), Xray diffraction (XRD), UV-Vis, and PL spectroscopy. The enhancement and suppression results can be fitted excellently with the modified Stern-Volmer equation, and the detection of arsenic is possible using these linear fit equations as calibration curves.

Key Words: Dichalcogenides; $MoS_2$; Quantum Dots; Arsenic; Photoluminescence

# 1. Introduction

Heavy metals are a source of groundwater contamination all over the world. Among the heavy metals, arsenic (As) is toxic and hazardous to human health [1]. The sources of arsenic can be either organic or inorganic. Inorganic sources are known to be more responsible for contamination than organic sources. Within the inorganic framework, the As (III) oxidative form is more dangerous, as it absorbs 50 times more and is less detectable compared to the As (V) oxidative form. The window for As (III) contamination is also known to be maximum at room temperature at a pH close to the neutral value [2]. Considering the adverse effect of arsenic on human health, the World Health Organization suggested the permissible limit of arsenic in drinking water to 10 ppb [3]. The detection of such a lower concentration of arsenic in aqueous media is challenging. Instead of very costly, time- and resource consuming, trained human resource-dependent analytical methods [4-6], it has become a genuine research challenge to develop easy and cheap methods that are quicker. The optical sensing method based on fluorescence techniques is one such focus area. In recent years, two-dimensional (2D) transition metal dichalcogenides (TMDs) have received increasing attention for various applications [7]. Molybdenum disulfide ($MoS_2$) is one such TMDs. When the size of the $MoS_2$ is reduced to an ultrathin 2D-structure or zero dimensional structure (quantum dots), it features excellent mechanical and optoelectronic properties [8, 9]. The nanosized $MoS_2$ provides versatile chemical properties and very good dispersion in aqueous media without any surfactant [10, 11]. $MoS_2$ QDs also showed strong excitation-dependent fluorescence, mainly because of the quantum confinement effect [12]. They are also known to have three different forms of sulfur, as determined from XPS studies [13]. The absorption and emission spectra as well as electrochemical redox potential changes with the change in particle size comparable to that of excitons (quantum confinement of conduction band electrons and valance band holes) [14, 15]. Additionally, surface states and trap levels are created in these types of quantum dots. Surface recombination and trapping are known to play significant roles in determining the luminescent properties of $MoS_2$ QDs [16].

The potential of $MoS_2$ as a material has been widely explored in various fields, including optoelectronic, energy storage, energy harvesting, catalysis, biomedicine, and sensing applications [17-20]. Several modalities, such as field-effect transistors (FET), electrochemical, surface plasmon resonance (SPR), electro-chemi-luminescence and photoluminescence (PL) have been considered for the development of a MoS2 based sensor [21, 22]. Among them, optical detection methods are more attractive because of their operational simplicity and high sensitivity.

Quantum dots (QDs) have been used for spectroscopic determination of biological as well as chemical molecules. Currently, QDs are utilized in a massive way to develop chemo-sensors for most toxic heavy metals. Generally, QDs are used both as passive and active fluorescent levels. The passive QD labels replace organic fluorophores for the detection of heavy metals. Active QD labels are dependent on charge transfer and various energy transfer mechanisms. $MoS_2$ QDs work as active labels.

Plenty of reports on PL-based detection of biomolecules and organic molecules exploiting the fluorescence property of $MoS_2$ QDs are available in the literature [11, 13, 23]. However, a limited number of reports are available for the detection of metal ions, such as $Ag^+$, $Pb^{2+}$, and $Fe^{+3}$ etc. [24-26]. In all cases, the $MoS_2$ QDs were functionalized with suitable functional molecules mainly to impart selectivity, and the QDs were used as a fluorescence quencher. Furthermore, an equal or higher than the permissible limit of the analyte concentration is also observed as the minimum test concentration in most of the cases. To the best of our knowledge, there is no report on the concentration-dependent PL study of bare $MoS_2$ QDs in the presence of As (III) in aqueous media. Therefore, an insight and comprehensive understanding of the PL properties of bare $MoS_2$ QDs in the presence of As (III) solution of wide concentration range are of great importance for further extension towards the development of new $MoS_2$ QD-based arsenic sensors. In this work, water-dispersed $MoS_2$ QDs were synthesized via solvothermal routes and the QDs were excited at a fixed wavelength of 360 nm, and the photoluminescence (PL) of excitons was studied over an extremely wide range of As (III) concentrations varying from 0.1 ppb to 1000 ppb. The lowest point of aqueous arsenic concentration was determined to find out the point at which this sensing is reasonable. The highest point of aqueous As (III) concentration was kept in the same order as the maximum level of aqueous As (III) concentration in groundwater [1]. The experimental results of the concentration-dependent PL showed an enhancement of PL of the $MoS_2$ QDs within the As (III) concentrations from 0.1 ppb to 50 ppb, followed by a suppression up to 1000 ppb. The concentration-dependent PL results also showed linear behavior and can be represented in two distinctive straight lines: one for the enhancement and the other for the suppression. This behavior of the QDs, concentration-dependent PL enhancement followed by suppression is unique, and to the best of our knowledge, has never been reported before.

## 2. Experimental section

Molybdenum disulfide ($MoS_2$), sodium arsenite (NaAsO2), N,N,-dimethyl formamide (DMF), poly-vinyl-pyrrolidone [PVP/(C)6H9NO)n] with molecular weight (MW) 58,000 and cetyl trimethyl ammonium bromide (CTAB) were procured from Alfa Aesar. Analytical grade chemicals were procured and used without purification. Deionized (DI) water from Milli-Q was used for the synthesis.

## 2.1. Preparation of MoS$_2$ quantum dots (QDs)

Bulk MoS$_2$ with CTAB was first exfoliated to nano-sheets by the solvent exfoliation method with slight modification, as reported by Coleman et al. [7]. Typically, 1 g each of MoS$_2$ and CTAB was added to 100 mL of DMF and kept under sonication for 10 h to exfoliate the MoS$_2$ powder. It was then centrifuged, and the supernatant was collected for further use. MoS$_2$ QDs were synthesized by a solvothermal approach, as suggested by Xu et al. with slight modifications [27]. Briefly, the supernatant of the MoS$_2$ sheets was refluxed at 180 0C for 8 hours followed by cooling at room temperature. It was centrifuged to separate the light yellow colored (**Figure 4.c (inset-i)**) supernatant and collected as MoS$_2$ QDs.

## 2.2. Sensing of arsenic by MoS2 QDs

A stock solution of 200 ppm Arsenic (III) was prepared by dissolving NaAsO$_2$ in DI water. Lower concentrations of arsenic and MoS$_2$ QDs were obtained by serial dilution using a buffer solution of pH 7. The resulting arsenic and MoS2 QDs solutions were mixed in 1:3 ratios (As: MoS$_2$ QDs). Corresponding MoS2 QDs solutions containing 0.1 to 1000 ppb As (III) were kept for 15 mins before PL measurements. All PL spectra were obtained using a spectro-fluorophotometer (OLYMPUS IX71).

## 3. Results and discussion

## 3.1. Verification of the formation of MoS2 QDs

MoS$_2$ QDs were synthesized from MoS$_2$ nanosheets. **Figure 1** shows the UV–Vis spectra of the as-prepared MoS$_2$ nanosheets and QDs. MoS$_2$ nanosheets are known to show four distinct excitonic peaks, which are generally allocated as A (667 nm), B (608 nm), C (443 nm), and D

(394 nm) excitons [20]. The first two peaks A and B are doublet peaks, which are the excitonic interband transition of 2D $MoS_2$ at the K point of the Brillouin zone, due to the splitting of the spin-orbit transition at the K point [28, 29]. The other two peaks, C and D, originate from the interband evolution between occupied and unoccupied orbitals [32].

However, in $MoS_2$ QDs, all these peaks disappear, except for a peak near the UV region ($\lambda < 300$ nm), resembling the excitonic features of $MoS_2$ QDs [30, 31]. The broad absorption at 318 nm seen in $MoS_2$ QDs is due to the convolution of all the excitonic peaks [32], which is blue-shifted as a result of the quantum size effect [20]. The crystallographic structure of the as-synthesized QDs was recorded using powder XRD, as shown in **Figure 2**. The observed diffraction peaks for $MoS_2$ QDs are at $2\theta = 14.53$, 39.91, and 49.34, corresponding to the (002), (103), and (105) planes, respectively, indicating the formation of mono- or few-layered $MoS_2$ QDs [33].

The TEM/HRTEM micrographs depicted in **Figure 3** represent the microstructures and morphologies of the as-synthesized $MoS_2$ QDs. The presence of almost spherical and well dispersed QDs is evident in **Figure 3.a**. A histogram was generated based on the information available from the observed 95 QDs and is presented in **Figure 3.b**. The average size of the QDs was found to be ~4.5 nm. Lattice fringes of $MoS_2$ QDs are visible in the HRTEM image depicted in **Figure 3.c**. The calculated fringes from the HRTEM image were found to be 0.21 nm [34], which corresponds to the (006) inter planar spacing of MoS2. **Figure 3.d** depicts the TEM image of the $MoS_2$ QDs after As (III) adsorption.

### 3.2. Spectral analysis of MoS2 QDs and their interaction with arsenic

The optical properties of the QDs were investigated using UV-Vis absorbance and photoluminescence (PL) measurements. When the aqueous $MoS_2$ QDs were treated with standard As (III) solutions with varied concentrations from 0.1 ppb to 50 ppb, an enhancement in absorbance was observed with the increase in arsenic concentration **(Figure 4.a)**. However, the opposite phenomenon, a constant decrease in absorbance with increasing arsenic concentration was observed when the QDs were treated with solutions of 50 ppb to 1000 ppb of As (III) **(Figure 4.b)**.

The as-prepared QDs were luminescent **(Figure 4.c (inset-ii))** and also showed concentration-dependent PL spectra. Hence, these QDs can be employed as a PL probe for the detection of arsenic in aqueous media. **Figure 4.c** and **Figure 4.d** shows the PL spectra of the QDs incorporated with varying concentrations of As (III) from 0.1 to 50 ppb and 50 ppb to 1000 ppb, respectively, in water. With increasing concentration, an enhancement in PL intensity up to 50 ppb and suppression beyond 50 ppb of As (III) concentration was observed when the QDs were excited with a constant laser operating at 360 nm. These trends of As (III) concentration-dependent PL measurements are similar to the previously recorded UVVis spectra under similar conditions.

Initially, the adsorption of As (III) happens at a low concentration on the surface of the $MoS_2$. The luminescence of these QDs is very sensitive to their surface states, which may change due to physical or chemical interactions, leading to a change in the efficiency of radiative recombination. In the case of suppression, various phenomena such as inner-filter effects, non-radiative recombination pathways, and electron transfer processes may be the reason [35]. In our case, initially, we see an enhancement of the photoluminescence of $MoS_2$ QDs in the presence of aqueous As (III), which is a rare phenomenon [36, 37]. Passivation of trap states or defects may be the reason for such an enhancement at relatively low concentrations of aqueous As (III). However, when the concentration of aqueous As (III) is beyond a particular limit, which in this case is 50 ppb, the extremely small glassy $As_2S_3$ particles may form on the $MoS_2$ QD surface.

The aqueous oxidative state of As (III) may then interact with these As3+ rich sites and create a 3d10-3d10 metallophilic interaction leading to new energy accepting levels on the As3+, which facilitates the Dexter energy transfer to As$^{3+}$ and suppression of the initial enhancement [38, 39]. The Stern-Volmer equation is usually applied to fluorescence quenching measurements, but a modified equation derived from the original Stern-Volmer equation can be applied to fluorescence enhancement measurements. The trends of fluorescence were analyzed through the utilization of this modified Stern-Volmer equation [40], which is written as:

$$\frac{F_0}{F - F_0} = \frac{1}{K_{SV}[C]} + 1$$

where $F_0$ and $F$ represent the PL intensities of MoS$_2$ QDs in the absence and presence of arsenic, respectively, [C] is the total arsenic concentration, and $K_{SV}$ is the modified Stern- Volmer constant.

Although the fluorescence enhancement up to 50 ppb As (III) was followed by gradual suppression of a further increase in As (III) concentration, we used the same equation to fit enhancement and suppression trends. Concentration-dependent PL results in the form of a modified Stern-Volmer equation are plotted in **Figure 5**. The correlation coefficient (R2) values for the enhancement **(Figure 5.a)** and suppression **(Figure 5.b)** trends are very close to 1, which proves that the data is well-fitted by the modified Stern-Volmer equation. It is noteworthy to mention that since the same equation applicable to fluorescence enhancement was also applied to quantify suppression, the corresponding slope of the suppression curve was negative, where the minus sign mathematically denoted the suppression in fluorescence. Also the K$_{SV}$ for the suppression trend had a negative sign because K$_{SV}$ is calculated as the reciprocal of the slope. These two trend lines can be used as calibration curves for the detection of arsenic. The arsenic concentration of an unknown sample can be predicted using the calibration curves in **Figure 5.a** or **Figure 5.b**. The selection of the right calibration can be decided based on the information of the PL measurements of the test samples, with and without dilution.

## 4. Conclusions

In summary, we have successfully demonstrated the synthesis of highly luminescent $MoS_2$ QDs through a solvothermal route and their use as fluorescent probes. These QDs are extremely sensitive and easily detect aqueous As (III) within the concentration range of 0.1−1000 ppb. The photoluminescence of these QDs was found to increase from 0.1 ppb to 50 ppb, and then gets suppressed till 1000 ppb. The QDs can detect arsenic concentrations as low as 0.1 ppb, which is well below the WHO permissible limit of arsenic in drinking water. $MoS_2$ QDs will prove to be a promising optical probe for detecting the presence of the As (III) oxidative state in water. The optical absorbance of these QDs was found to follow a similar pattern.

**Acknowledgments**

**JRA** gratefully acknowledges the support from GGSIPU, New Delhi in the form of a short term research fellowship. The authors are grateful to the AIRF, JNU, New Delhi, India for carrying out TEM characterizations. **AD** is thankful to GGSIPU, New Delhi for financial support under the FRGS grant [GGSIPU/DRC/FRGS/2018/1(1115)] and to DST for FIST grant [SR/FST/PSI-167/2011(C)].

**Author contributions**

**AD and TS** consived the experiments. **JRA** and **FN** grew, characterized the materials and performed the PL measurements. **NS** performed the UV-Vis measurements. All authors discussed the data and wrote the manuscript together.

**Competing financial interests**

The authors declare no competing financial interests.

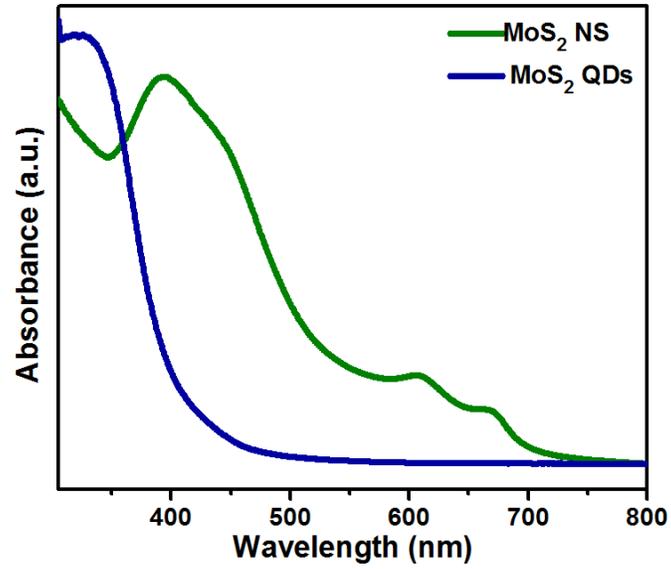

**Figure 1**: UV-Vis absorption spectra of MoS2 nanosheets (olive) and MoS2 QDs (navy).

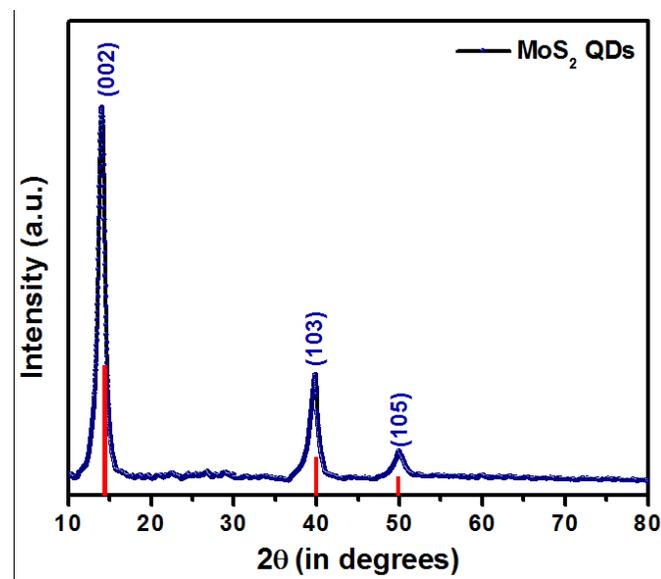

**Figure 2**: XRD pattern of MoS2 QDs. The red lines present the standard MoS2 (JCPDS No. 37-1492) peaks.

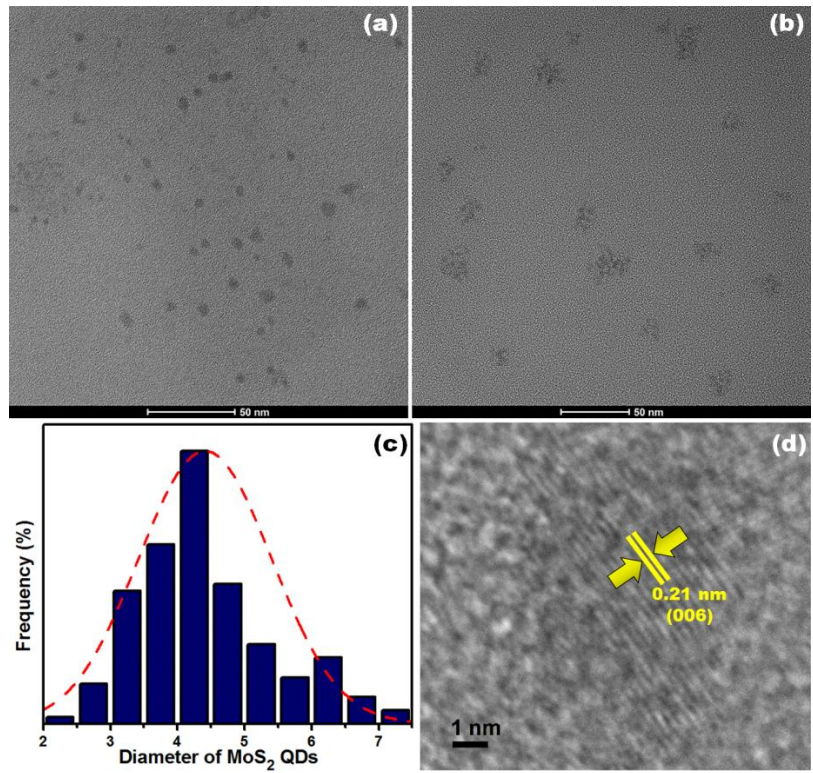

**Figure 3**: **(a)** TEM image of MoS2, **(b)** Histogram of MoS2 QDs size distribution, **(c)** HRTEM image of MoS2 QDs showing lattice spacing of 0.21 nm along the (006) planes, and **(d)** TEM micrograph of MoS2 QDs after arsenic adsorption.

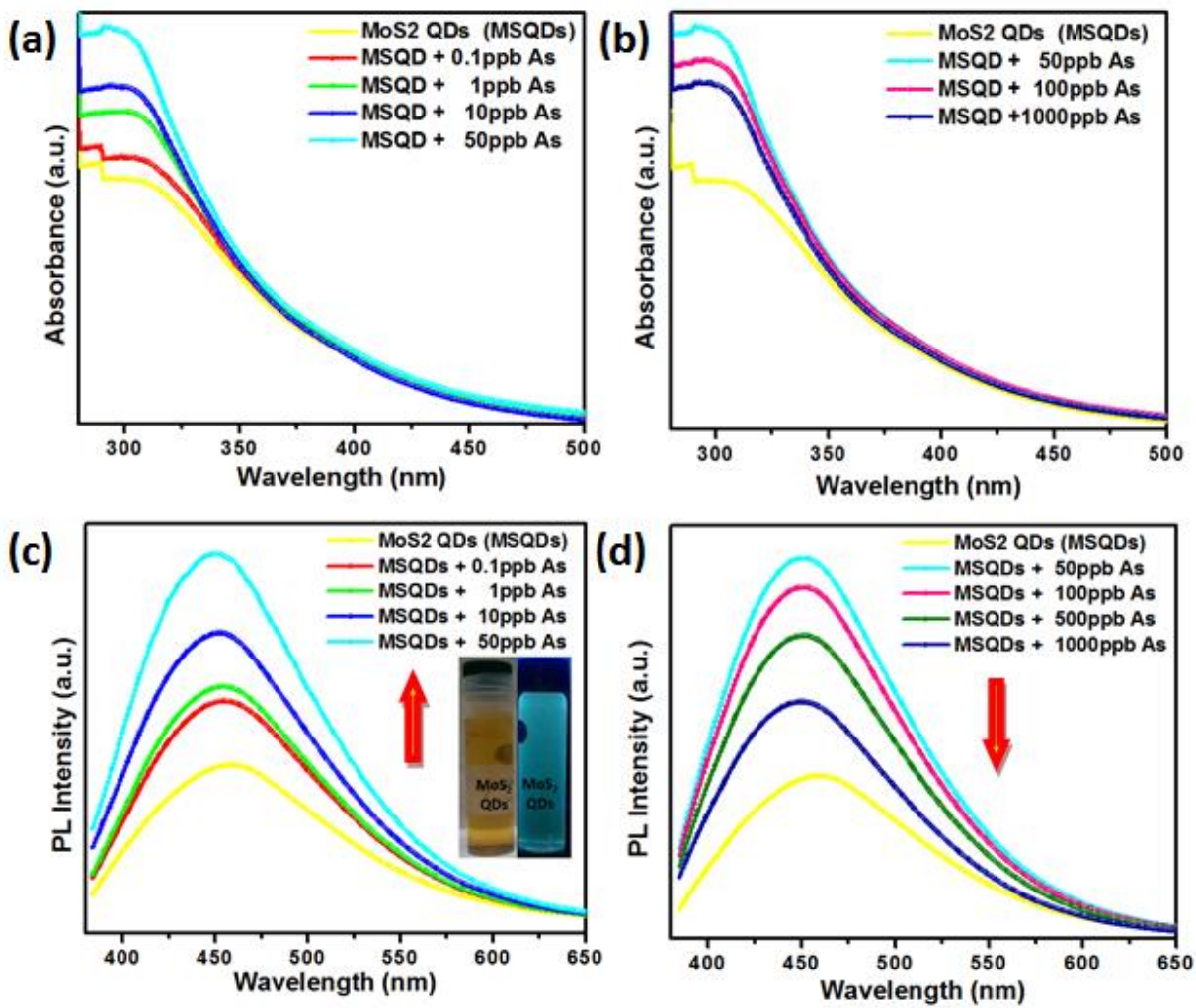

**Figure 4**: UV-Vis absorption spectra of MoS2 QDs in the presence of varying concentrations of As (III) from **(a)** 0 ppb to 50 ppb and **(b)** 50 ppb to 1000 ppb. Photoluminescence spectra of MoS2 QDs in the presence of varying concentrations of As (III) from **(c)** 0 ppb to 50 ppb and **(d)** 50 ppb to 1000 ppb. The **inset (Figure c)** represent photographs of MoS2 QDs **(i)** without UV irradiation and **(ii)** UV irradiation.

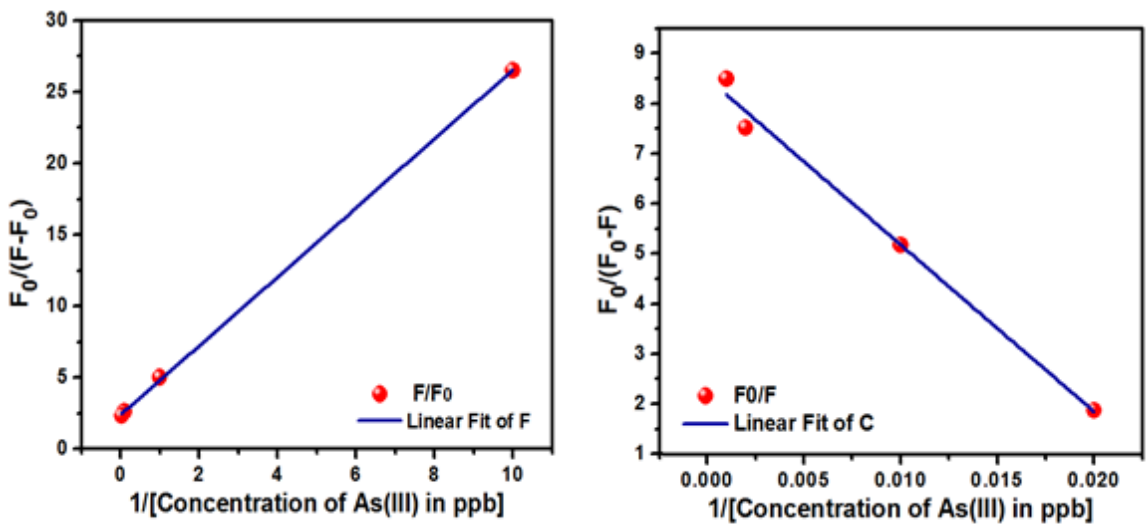

**Figure 5**: Concentration dependent modified Stern-Volmer plot within the concentration range of **(a)** 0.1 ppb to 50 ppb ($R^2$ = 0.99 and $K_{SV}$ = ~ 0.41 ppb-1), and **(b)** 50 ppb to 1000 ppb ($R^2$ = 0.99 and $K_{SV}$ = ~ -3.08 × 10-3 ppb-1).

**Graphical Abstract**

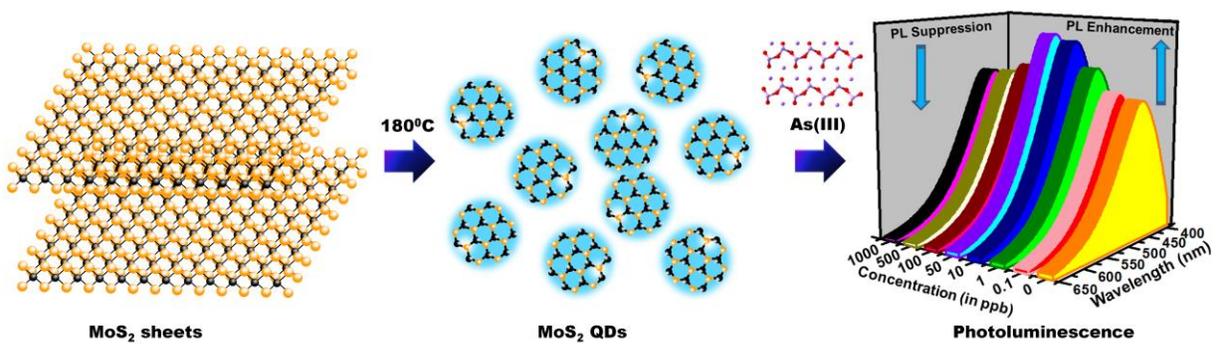